\documentclass[usenatbib]{mn2e}
\usepackage{amsmath, epsfig, natbib}

\begin{document}
\title [Effect of dark matter halo on global modes] 
{Effect of dark matter halo on global spiral modes in galaxies}
\author[S. Ghosh T. D. Saini and C.J. Jog ]
       {Soumavo Ghosh$^{1}$\thanks{E-mail : soumavo@physics.iisc.ernet.in}, 
 Tarun Deep Saini$^{1}$\thanks{E-mail : tarun@physics.iisc.ernet.in}, and
        Chanda J. Jog$^{1}$\thanks{E-mail : cjjog@physics.iisc.ernet.in}\\
$^1$   Department of Physics,
Indian Institute of Science, Bangalore 560012, India \\
} 
\maketitle

\begin{abstract} 
Low surface brightness (LSB) galaxies form a major class of galaxies, and are 
characterized by low disc surface density and low star formation rate. These 
are known to be dominated by dark matter halo from the innermost regions. Here 
we study the role of dark matter halo on the grand-design, $m=2$, spiral modes 
in a galactic disc by carrying out a global mode analysis in the WKB approximation. 
The Bohr-Sommerfeld quantization rule is used to determine how many discrete global 
spiral modes are permitted. First a typical superthin LSB galaxy, UGC 7321 is 
studied by taking only the galactic disc, modelled as fluid; and then the disc 
embedded in a dark matter halo. We find that  both cases permit the existence of 
global spiral modes.  This is in contrast to earlier results where the 
inclusion of dark matter halo was shown to nearly fully suppress local, 
swing-amplified spiral features. Although technically global modes are permitted in the fluid model
as shown here, we argue that due to lack of tidal interactions, these are not 
triggered in LSB galaxies. For comparison, we carried out a similar analysis for 
the Galaxy, for which the dark matter halo does not dominate in the inner regions. 
We show that here too the dark matter halo has little effect, hence the disc 
embedded in a halo is also able to support global modes. The derived pattern speed 
of the global mode agrees fairly well with the observed value for the Galaxy. 
\end{abstract}

\begin{keywords}
{galaxies: kinematics and dynamics - 
 galaxies: spiral - galaxies: structure - galaxies: individual: UGC 7321 - galaxies: haloes - instabilities}
\end{keywords}

\section{Introduction} 
Various surveys on galaxy morphology have revealed that the spiral arms, making a spectacular visual impression, are mainly of two types, namely, grand-design spiral arms and flocculent arms. It is also found that the fractional abundance of these two types varies with the Hubble type \citep{Elm11}. The origin and maintenance of these features has been extensively studied over the past five decades, though many aspects are still not fully understood.
In general, the grand-design spiral features are explained as density waves, governed mainly by gravity \citep{LS64, LS66}, whereas the flocculent spiral features are material arms, caused by swing amplification \citep{GB65, Tom81}.

A major class of galaxies are the so-called low surface brightness (LSB) galaxies, which are characterized by low star formation rate \citep{IB97}, and low disc surface density \citep{dBM96,dBM01}.
LSB galaxies do not show grand-design spiral structure. These can show spiral structure but it is fragmentary,  extremely faint, and difficult to trace \citep{Sch90, Mcg95,Sch11} The LSBs are dark matter dominated starting right from the very inner radii
\citep{Bot97,dBM97,dBM01,Com02,Ban10}. This aspect of dark matter dominance is different from that seen in case of normal spiral galaxies. For example, in the LSBs, the dark matter constitutes about 90 percent of the total mass within the optical disc, whereas for the 'normal' or High Surface Brightness (HSB) galaxies, such as the Milky Way, the contribution of stellar mass and dark matter halo mass is comparable \citep[e.g.][]{dBM01, Jog12} within the optical disc. In the Milky Way and other HSB galaxies, the dark matter dominates only in regions way outside the optical disc. Consequently, the LSB galaxies naturally offer a good place for probing the effect of the dark matter halo on the dynamics  of a galactic disc. We caution that we only consider the small size LSBs which are more common, and do not discuss giant LSBs like Malin1. The latter sub-class have massive discs and can show fairly strong spiral arms as in UGC 6614 \citep{Das13}.

In the literature, several studies have demonstrated that the dark matter halo has a profound effect on various dynamical properties of LSBs. For example, early numerical simulation by \citet{Mih97} showed that the dominant dark matter halo and the low surface density make the LSB galaxies stable against the growth of global non-axisymmetric modes such as bars. Further, there exists a subclass of LSBs, namely, superthin galaxies. A study by \citet{BJ13} showed that the superthin property is explained by a dense, compact halo that dominates from the innermost regions.
 
A recent work by \citet{GJ14} showed that a dark matter halo that is dominant from the innermost regions  makes the galactic disc stable against both local axisymmetric and non-axisymmetric perturbations, by making the Toomre Q parameter very high ($>$ 3). This suppresses the swing amplification process, thereby explaining the observed lack of star formation and lack of strong small-scale spiral features in LSB galaxies. But, the role of dark matter has not been studied so far in the context of grand-design spiral structure. Motivated by the results for small-scale spiral structure, here we plan to study the effect
 of dark matter halo in the context of global spiral modes.

The idea that spiral patterns could be long-lived density waves was first proposed by \citet{LS64}. In this hypothesis the spiral pattern is a rigidly rotating density wave through which differentially-rotating stars and gas can flow. The pattern is maintained by the self-gravity and pressure of the density wave self-consistently. The pattern retains its shape over long periods of time without suffering the winding problem. Investigation of self-sustaining density waves in a disc is essentially a problem of wave-mechanics in a differentially rotating, self-gravitating disc. However, the long-range nature of gravitational interaction makes studying them difficult due to the non-local gravitational interactions between different parts of the perturbed disc. 

A very useful approach for studying the dynamics of discs is the tight-winding approximation. This approximation makes the gravitational effects of the density perturbation local. For a wave-like, non-axisymmetric, small amplitude perturbation, the temporal oscillation frequency $\omega$ can be obtained as a function of the radial wavevector $k$, the radial coordinate $R$, and the azimuthal wavenumber $m$, known as WKB dispersion relation \citep[see, e.g., Chapter~6 of][for derivation and applications]{BT87}. The same local dispersion relation can also be used to construct global standing-wave like solutions through the Bohr-Sommerfeld quantization condition. In the past this approach has been used by \citet{BR88}, \citet{Shu90}; and more recently by \citet{Tre01}, \citet{Sai09}, \citet{Gul12} for $m=1$ modes. Here we are using this alternative approach to investigate the role of dark matter on the existence of large-scale spiral patterns in a disc galaxy.

In this paper we investigate the number of allowed discrete global mode(s)---the term is used as a proxy for the grand spiral arms---within the WKB approximation, using the Bohr-Sommerfeld quantization condition. We treat the stellar disc as a fluid, which allows a considerably simple analytical dispersion relation. This assumption is valid so long as one is away from the resonance points (although see \S~4.1). However, we remove the artefacts, intrinsically associated with the fluid approximation where necessary (for details see \S~3.1). Our analysis is carried out for UGC 7321, a superthin, LSB galaxy, first for a stellar disc alone and then for a disc embedded in a dark matter halo. We find that in both the disc-alone and disc plus halo models, global spiral modes are present i.e. based on this linear calculation and under fluid approximation, dark matter does not produce any significant change for global spiral modes. Similarly, we followed the same course of modal analysis for  our Galaxy, a typical 'normal', HSB galaxy. For our Galaxy, dark matter halo is shown to have negligible effect on global modes. This trend for our Galaxy follows our expectation since dark matter halo is not a dominant component in the inner/optical region of the Galaxy. For this work  the observed profiles of surface density and stellar velocity dispersion for both UGC 7321 and the Galaxy are used, so as to make our calculation more realistic.

In \S~2  we present the model used, and the details of model input parameters; \S~3 presents the details of the WKB approximation and \S~4 describes the results while \S~5 and \S~6 contain the discussion and conclusions respectively.

\section{Formulation of the Problem}
We treat the galactic stellar disc as a fluid, characterized by an exponential surface density $\Sigma_s$, and the one-dimensional velocity dispersion $v_s$ for pressure support, which is treated as the fluid sound speed. To keep our analysis simple the disc is taken to be infinitesimally thin. The disc is embedded in a spherically symmetric dark matter halo with a pseudo-isothermal radial profile. We have used the cylindrical coordinates ($R$, $\phi$, $z$) for both disc as well as the spherical halo. The dynamics of the disc is calculated first under the gravity of disc only (referred to as the disc-alone case) and then under the combined gravity of the halo and the disc (referred to as the disc plus dark matter case). The halo is assumed to be gravitationally inert, i.e.,  it is assumed to be non-responsive to the gravitational force of the perturbations in the disc.  

\subsection{Model of disc and halo}
\label{sec:model}
For a galactic disc embedded in a dark matter halo concentric to the galactic disc, the net angular speed, $\Omega$, and the net epicyclic frequency, $\kappa$ (used later in the dispersion relation), are obtained by adding the disc and the halo contribution in quadrature as follows:
\begin{equation}
\kappa^2= \kappa^2_{\rm disc}+\kappa^2_{\rm DM}\,; \quad \Omega^2= \Omega^2_{\rm disc}+\Omega^2_{\rm DM}\,.
\end{equation}
The frequencies can be computed from the standard expressions in terms of the gravitational  potentials of the disc and the halo. For an exponential disc, the potential $\Phi(R, 0)$ in the equatorial plane is given by \citep[][equation (2.168)]{BT87}
\begin{equation}
\Phi(R, 0)= -\pi G \Sigma_0R [I_0(y)K_1(y) - I_1(y)K_0(y)]\,,
\end{equation}
where $\Sigma_0$ is the disc central surface density, $y$ is the dimensionless quantity, defined as $y = R/2R_d$, where $R$ is the galactocentric radius, $R_d$ is the exponential disc scale-length. $I_n$ and $K_n$ $(n= 0, 1)$ are the modified Bessel functions of first and second kind, respectively. In terms of the disc potential,  $\kappa^2_{\rm disc}$ and $\Omega^2_{\rm disc}$ are given as
\begin{equation}
\begin{split}
 \kappa^2_{\rm disc} = \frac{\pi G \Sigma_0}{R_d}  \Big[4I_0(y) & K_0(y)  -2I_1(y) K_1(y) +\\
   & 2y\left(I_1(y) K_0(y) - I_0(y) K_1(y) \right)\Big]
\end{split}
\end{equation}
and
\begin{equation}
\Omega^2_{\rm disc}= \frac{\pi G \Sigma_0}{R_d}\Big[I_0(y) K_0(y) - I_1(y) K_1(y) \Big]\,.
\end{equation}

For both the galaxies considered in this paper, UGC 7321 and the Galaxy, a pseudo-isothermal profile has been shown to yield a good fit for the DM halo \citep{Mer98, Ban10}. The gravitational potential of a pseudo-isothermal halo is given by
\begin{equation}
\begin{split}
\Phi_{\rm DM}=4 \pi G \rho_0 R^2_c \Bigg[\frac{1}{2}\log (R^2_c+R^2+z^2)+\\
\quad\Bigg(\frac{R_c}{(R^2+z^2)^{1/2}}\Bigg)\times \tan^{-1}\Bigg(\frac{(R^2+z^2)^{1/2}}{R_c}\Bigg)-1\Bigg]\,,
\end{split}
\end{equation}
where $R_c$ is the core radius and $\rho_0$ is the core density. The corresponding  $\kappa^2_{\rm DM}$ and $\Omega^2_{\rm DM}$, in the mid-plane ($z$ = 0), are given as
\begin{equation}
\begin{split}
\kappa^2_{\rm DM}=4 \pi G \rho_0 \Bigg[\frac{2}{1+(R/R_c)^2}+\Big(\frac{R_c}{R}\Big)^2\frac{1}{1+(R/R_c)^2}-\\
\Big(\frac{R_c}{R}\Big)^3\tan^{-1}\Big(\frac{R}{R_c}\Big)\Bigg]\,,
\end{split}
\end{equation}
and
\begin{equation}
\begin{split}
\Omega^2_{\rm DM}=4 \pi G \rho_0\Bigg[\frac{1}{1+(R/R_c)^2}+\Big(\frac{R_c}{R}\Big)^2\frac{1}{1+(R/R_c)^2}\\
-\Big(\frac{R_c}{R}\Big)^3 \tan^{-1}\Big(\frac{R}{R_c}\Big)\Bigg]\,.
\end{split}
\end{equation}

\subsection{Model parameters}
To calculate the rotation and epicyclic frequencies needed for the WKB analysis below, we have used the following set of parameters.
\subsubsection{UGC 7321}
For the stellar disc and halo parameters, the values we used are obtained either observationally or by modelling (Banerjee et al. 2010). The stellar disc is an exponential disc with central surface density of 50.2 M$_\odot$ pc$^{-2}$ and a disc scale length of 2.1 kpc. For the dark matter halo, a pseudo-isothermal profile with core density of 0.057 M$_\odot$ pc$^{-3}$ and core radius of 2.5 kpc yields good fit for the observed rotation curve and the vertical gas scale-height distribution \citep{Ban10}.

For the one-dimensional stellar velocity dispersion in the radial direction, the profile is taken as $v_s$ =$v_{s0} \exp(-R/2R_d)$. The observed central dispersion along $z$ is equal to 14.3 km s$^{-1}$ \citep{Ban10}. For the solar neighbourhood, it is observationally found that, the ratio of velocity dispersion in the $z$ direction to that of radial direction is $\sim$ 0.5 \citep[e.g.,][]{BT87}. Here we assume the same conversion factor for all radii in this galaxy.

\subsubsection{The Galaxy}
The stellar disc parameters for the Galaxy are taken from the standard mass model by \citet{Mer98} which gives an exponential radial profile for the disc surface density with a disc scale-length of 3.2 kpc and a central surface density of 640.9 M$_\odot$ pc$^{-2}$. For the dark matter halo, a pseudo-isothermal profile is used with a core radius of 5.0 kpc and a core density of 0.035 M$_\odot$ pc$^{-3}$ \citep{Mer98}.

The one-dimensional stellar velocity dispersion is observed to be of  the form:
$v_s = v_{s0}\exp(-R/8.7)$, where $v_{s0} = 95$ km s$^{-1}$ \citep{LF89}.

\section{WKB analysis}
\label{sec:WKB}

\begin{figure}
\centering
\includegraphics[height=2.5in,width=3.5in]{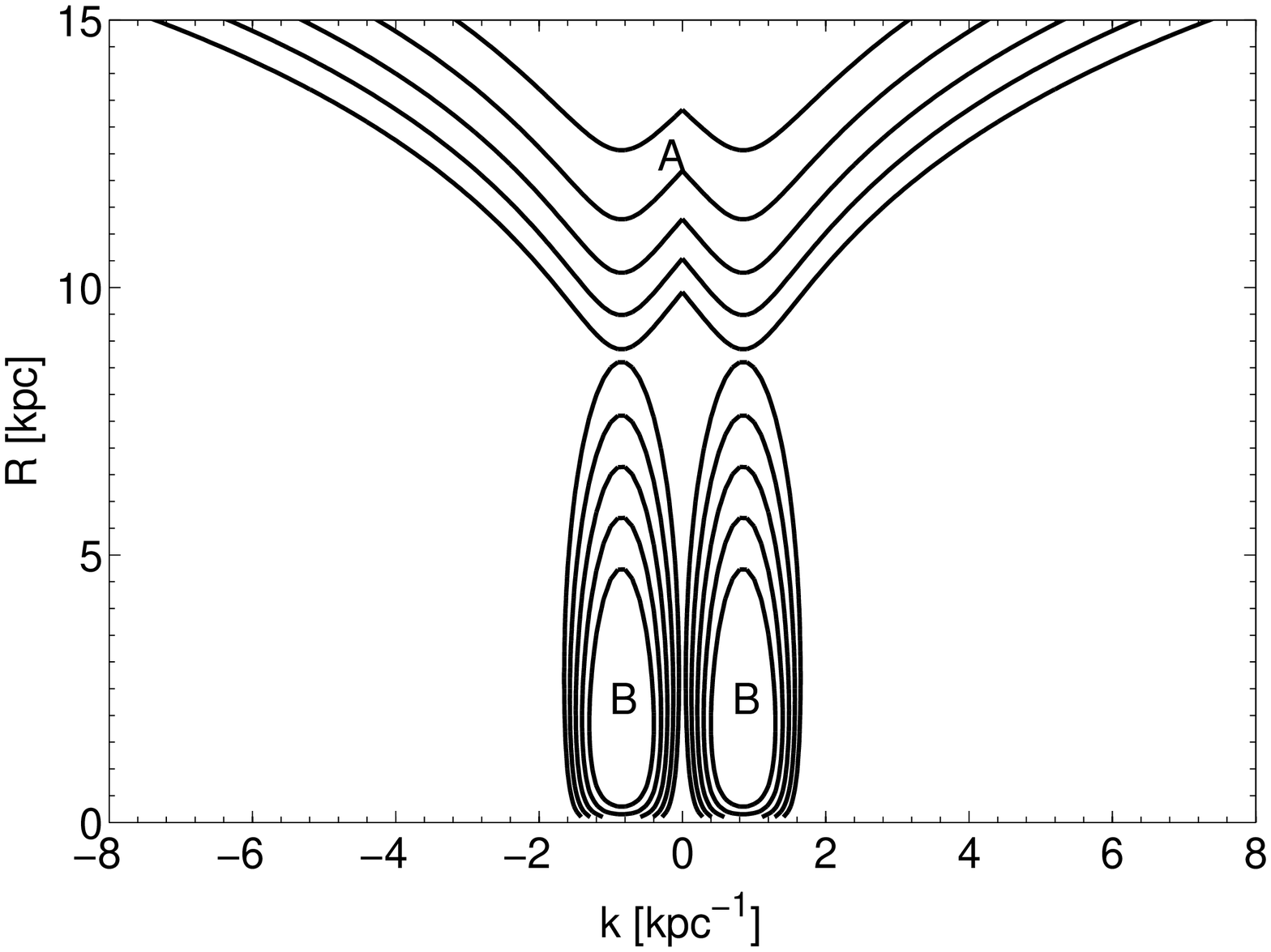}
\medskip
\includegraphics[height=2.5in,width=3.5in]{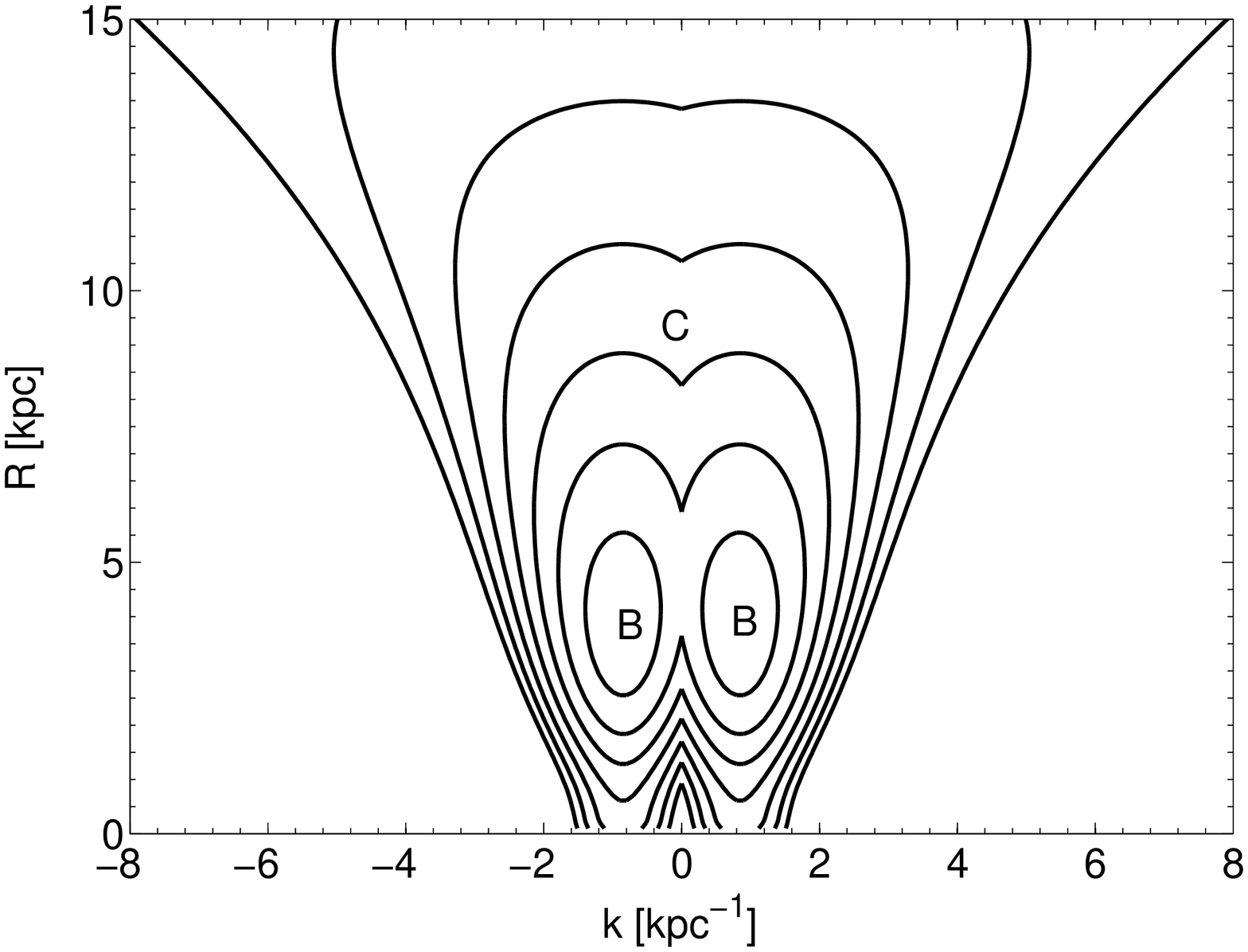}
\caption{Propagation diagrams (contours of constant $\Omega_{\rm p}$) plotted for different pattern speeds, by using the input parameters of LSB galaxy UGC 7321. The top panel shows contours for the disc-alone case where the range of $\Omega_{\rm p}$ varies from 2.3 km s$^{-1}$ kpc$^{-1}$ to 4.3 km s$^{-1}$ kpc$^{-1}$  and the bottom panel shows the contours for the disc plus dark matter halo case where the range of $\Omega_{\rm p}$ varies from 1.5 km s$^{-1}$ kpc$^{-1}$ to 4.1 km s$^{-1}$ kpc$^{-1}$. The contours are plotted at intervals of 0.2 for both the panels. Different types of contours present here are marked as A, B and C. $\Omega_{\rm p}$ value increases from the outer to the inner contours.}
\label{fig:UGC7321}
\end{figure}

For small non-axisymmetric perturbations of the form 
\begin{equation}
X(R,\phi, t) = X_a \exp \left [{\rm i} \left(\int^R k(R')dR' -m\phi + \omega t \right) \right]\,,
\end{equation}
where $X$ is a generic dynamical quantity, and assuming the tight-winding approximation, which formally requires $|kR| \gg 1$, the Euler equations reduce to the standard dispersion relation \citep{BT87}
\begin{equation}
(\omega - m\Omega)^2=\kappa^2-2\pi G \Sigma_s |k|+v^2_sk^2\,,
\label{disp-equation}
\end{equation}
where $\kappa$ and $\Omega$ are as defined in the \S~\ref{sec:model}, with the parameters given in the previous subsection. We are interested in spiral patterns corresponding to $m =2$. A slight rearrangement of equation (\ref{disp-equation}) yields
\begin{equation}
4(\Omega_p-\Omega)^2=\kappa^2-2\pi G \Sigma_s|k|+v^2_sk^2\,,
\end{equation}
where $\Omega_p= \omega/2$ is the pattern speed at which the spiral pattern rotates rigidly. For a given pattern speed $\Omega_p$, this dispersion relation can be thought of as an implicit relation between the phase space variables $k$ and $R$ (the resulting graph is also known as the propagation diagram).
 In general this relation is multi-valued, therefore, it is convenient to obtain its two branches explicitly for computational purposes. Expressing the relation in terms of the wavevector $k$ as a function of $R$ we get
\begin{equation}
|k_{\pm}|(R)=[2\pi G \Sigma_{s} \pm \Delta^{1/2}]/2v^2_s\,,
\label{wavevector}
\end{equation}
where 
\begin{equation}
\Delta = [(2\pi G\Sigma_s)^2-4v^2_s\{\kappa^2-4(\Omega_p-\Omega)^2\}] \,.
\end{equation}
Note that the dependence on the wavevector is through its absolute value $|k|$; and therefore, these two branches in general correspond to four possible values of $k$ at any given radius $R$. However, if $\Delta^{1/2} > 2\pi G \Sigma_s$, then only the '+' branch is physical, and the total number of solutions for $k$ reduces to two. Waves cannot exist in a region where $\Delta < 0$, called the forbidden region, where there are no real solutions for $k$. From equation~(\ref{wavevector}) it can be seen that when $\Delta=0$, $k_{+}=k_{-}$, the two branches merge at that point, and we say that the wave reflects from trailing to leading branch and vise versa.
 The Lindblad resonances occur when $\Omega_p = \Omega \pm \kappa/2$. The dispersion relation then gives $k=0$ for the $k_{-}$ branch. The $k_{+}$ branch, however, can extend beyond the Lindblad radius for gas. This generic behaviour
is evident in the constant $\Omega_p$ plots in Fig~\ref{fig:UGC7321} where the plot is for the LSB galaxy UGC 7321 with and without the dark matter halo.

A constant $\Omega_p$ (or equivalently constant $\omega$) contour represents the trajectory of a wave packet which travels at a radial velocity given by the corresponding group velocity $v_g = d\omega/dk$  \citep{Tom69}.

Figure~\ref{fig:UGC7321} exhibits three different types of contours. Their physical significance in terms of propagation of a wave packet is as follows.

\begin{itemize}
\item{{\it Type A}:} These are the 'W' shaped contours in the top panel of Fig.~\ref{fig:UGC7321}. A wavepacket may start at a large radius (to the left of the curve) on the short leading branch and then travel inwards with a negative radial group velocity. It first reflects from the edge of the forbidden region at the lowest point of the curve, moves radially outwards and reaches the outer Lindblad radius (at $k=0$) from where it is reflected back into the long trailing branch, and gets reflected again from the edge of the forbidden region. The wavevector of perturbations comprising the wavepacket then continues to increase steadily. In a collisionless disc, the fate of such wavepackets would be to eventually dissipate at the Lindblad resonance by a process similar to Landau damping \citep{BT87}. 

\item{{\it Type B}:} These are the oval shaped closed curves and they always occur in pairs. A wavepacket would move in and out indefinitely by being reflected at the edge of the two forbidden regions, from short branch to relatively long branch and vice versa, this being true for both the leading as well as the trailing branch. Also note that the wavevector remains non-zero all along the closed curve.

\item{{\it Type C}:} Unlike {\it Type B} contours, this type of closed contours are not degenerate. They involve reflection of a wave packet from the leading to the trailing branch, and vice versa, by the edge of the forbidden region as well as the inner and outer Lindblad radius.
\end{itemize}

\subsection{Global standing waves: the quantization condition}
In contrast to the behaviour of wavepackets, the global modes of a disc are characterized by a single frequency $\omega$ (or $\Omega_p$) and extend over a large region of the disc. These essentially non-local perturbations are characterized by a wavefunction that extends across the disc, and which is expected to behave in a manner consistent with the boundary conditions across the disc. The density perturbation across the disc for these standing waves has a specific pattern shape (given by the wavefunction) and pattern speed $(\Omega_p = \omega/2)$. 

 In principle, the pattern speed $\Omega_p$ could be either positive or negative, implying a prograde or retrograde motion for the pattern with respect to the galactic rotation. However, we find that for the two galaxies considered by us, $\Omega - \kappa/2 > 0$; and collisionless discs do not support wave-like solutions for negative pattern speeds under this condition, although fluid discs do allow them. We have verified that the resulting propagation diagrams for negative $\Omega_p$ are not closed, therefore, in the rest of the paper we do not consider negative $\Omega_p$. All the global modes described below are thus prograde.

\begin{figure}
\centering
\includegraphics[height=2.5in,width=3.5in]{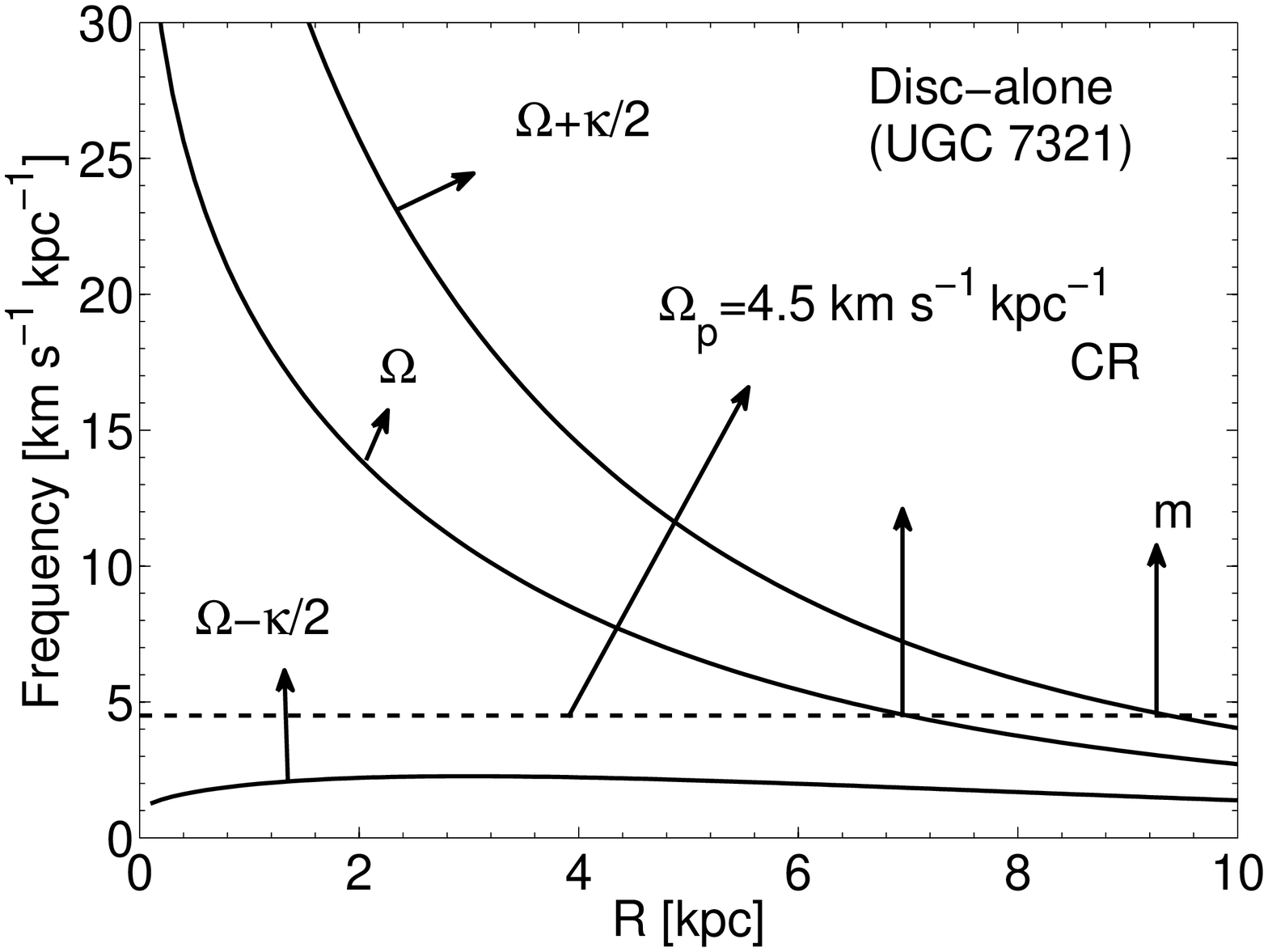}
\medskip
\includegraphics[height=2.5in,width=3.5in]{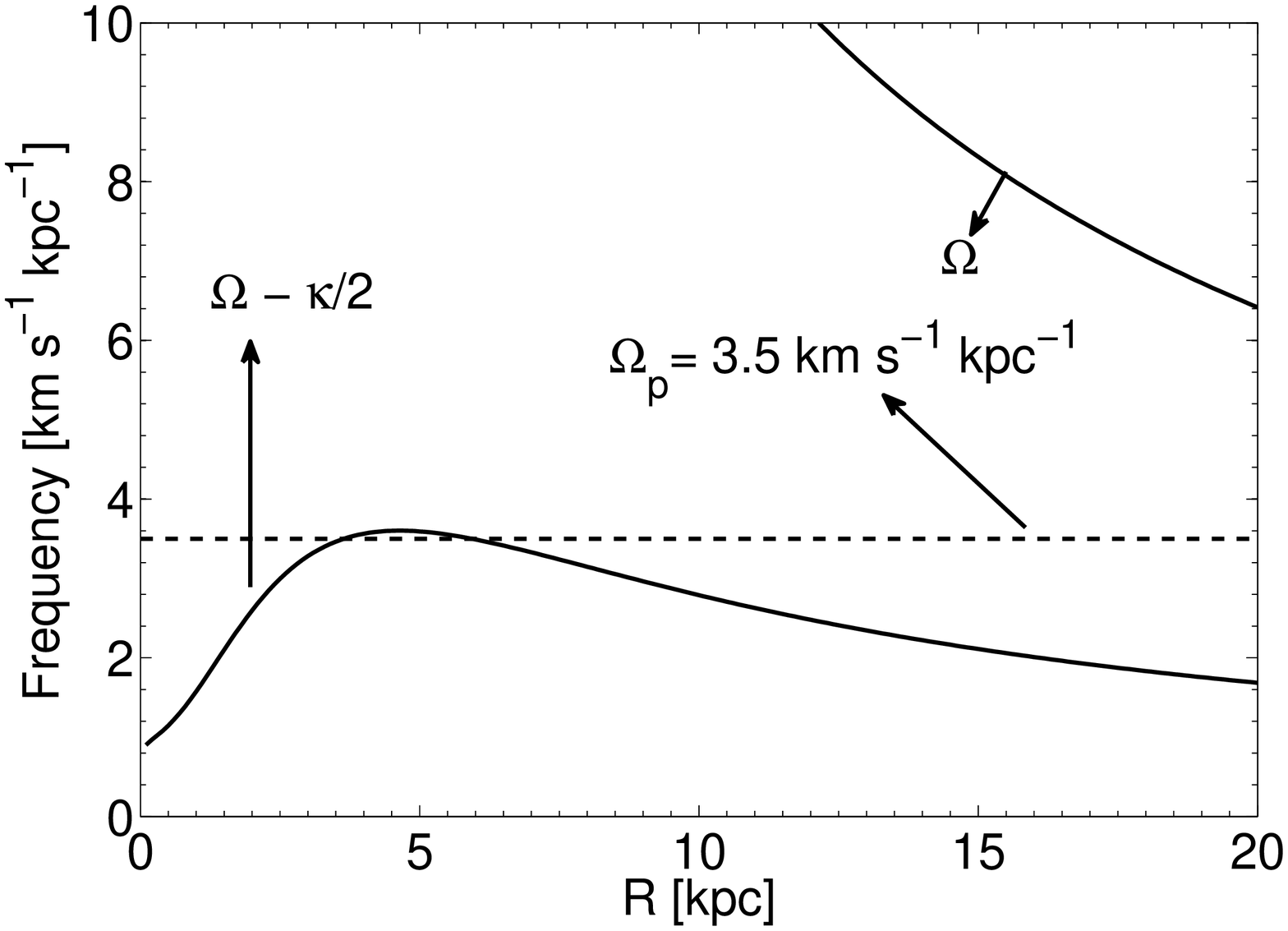}
\caption{ The $\Omega$, $\Omega-\kappa/2$ \& $\Omega+\kappa/2$ (in units of km s$^{-1}$ kpc$^{-1}$) curves are shown as a function of radius for UGC 7321. Top panel shows the disc-alone case, where the pattern speed ($\Omega_{\rm p}$) of 4.5 km s$^{-1}$ kpc$^{-1}$ gives a closed contour of {\it{Type B}} and the bottom panel showing disc plus dark matter halo case, where the pattern speed ($\Omega_{\rm p}$) of 3.5 km s$^{-1}$ kpc$^{-1}$ gives a closed contour of {\it{Type C}}. 
Note that, for {\it Type C} contour, the shown $\Omega_{\rm p}$ lies below $\Omega-\kappa/2$ for a certain range of $R$, thus it does not satisfy the inequality mentioned in the text (see \S ~3.1)
 }
\label{fig2}
\end{figure}

In the previous section we saw that the fate of {\it Type A} wavepackets is to eventually disperse, and unless the boundary condition at large radius is reflecting them back, they will not give rise to long-lived standing waves. However, the contours of {\it Type B \& C} seem to have the correct behaviour to give rise to global standing waves due to their closed shape. Interestingly, note that for contours of {\it Type~C}, the wave goes right past the Lindblad resonances on either side before getting reflected from the forbidden region. This behaviour is peculiar to fluid discs. For collisonless disc the waves can exist only in regions where $\Omega - \kappa/2 \le \Omega_p \le \Omega + \kappa/2$. In Fig.~\ref{fig2} we plot the line $\Omega = \Omega_p$ with $\Omega - \kappa/2$,  $\Omega$, and $\Omega + \kappa/2$, where we can see that the contours of {\it Type~B} do obey this condition, but not the contours of {\it Type~C}. 

To elaborate, in the fluid model a wave can partially transmit through the Inner Lindblad Resonance (ILR), and consequently the wave cycle, operating between the co-rotation (CR) and ILR to maintain a standing wave, will get hampered. In other words, to maintain a wave cycle between CR and ILR, one has to make sure that the inequality
\begin{equation}
 R_{ILR} < R_{-} < R_{+} < R_{CR}\,,
\label{ineq}
\end{equation} 
where $R_{\pm}$ are the boundaries of the forbidden region, holds for each mode, otherwise for that mode ILR would be exposed. Since the inner regions of galactic discs are dominated by stars, only the closed contours for which the above mentioned inequality holds can be realized in a real galaxy \citep[see page 205][for a detailed discussion]{Ber00}. Therefore, in our analysis the contours of {\it Type~C} are considered as unphysical, and hence are discarded.

To construct the discrete global density waves from the closed contours of {\it Type B}, it is useful to recall that in quantum mechanics the WKB approximation is used to calculate the eigenvalues of a Hamiltonian in the following manner: for a one dimensional system the phase plot (in the $p$--$q$ plane, where $q$ and $p$ are generalized coordinates and generalized momenta respectively) of constant energy for a bound state forms a closed loop.  The total area inside the loop $\oint pdq$ can then be shown to be some integral multiple of $\hbar$ \citep[see, e.g.,][]{LL65}. In our case this Bohr-Sommerfeld quantization rule essentially reduces to constructing the constant $\omega$ curves in the $k$--$R$ plane. Provided that closed loops exist, they can be quantized by calculating the $\oint kdR$ and equating it to integral multiples of $\pi$, apart from some extra factors of $\pi/2$ that arise due to the boundary condition at the turning points. The quantization condition ensures that only discrete values of $\omega$ are allowed, which are called the eigenfrequencies \citep[e.g.][]{BR88}.

For the closed contours of {\it Type~B} to represent standing waves they must satisfy a quantization condition. The appropriate WKB quantization rule is given by \citet{Tre01}
\begin{equation}
2 \int_{R_-}^{R_+} [k_{+}(R)-k_{-}(R)] dR = 2\pi (n-\frac{1}{2})\,
\label{qcond}
\end{equation}
where $n=1,2,3,\cdots$. Although the WKB approximation requires $|kR| \gg 1$, its validity is often sufficiently accurate even for $|kR| \simeq 1$. 

\section{Results}
\subsection{UGC 7321}
\begin{figure}
\centering
\includegraphics[height=2.5in,width=3.5in]{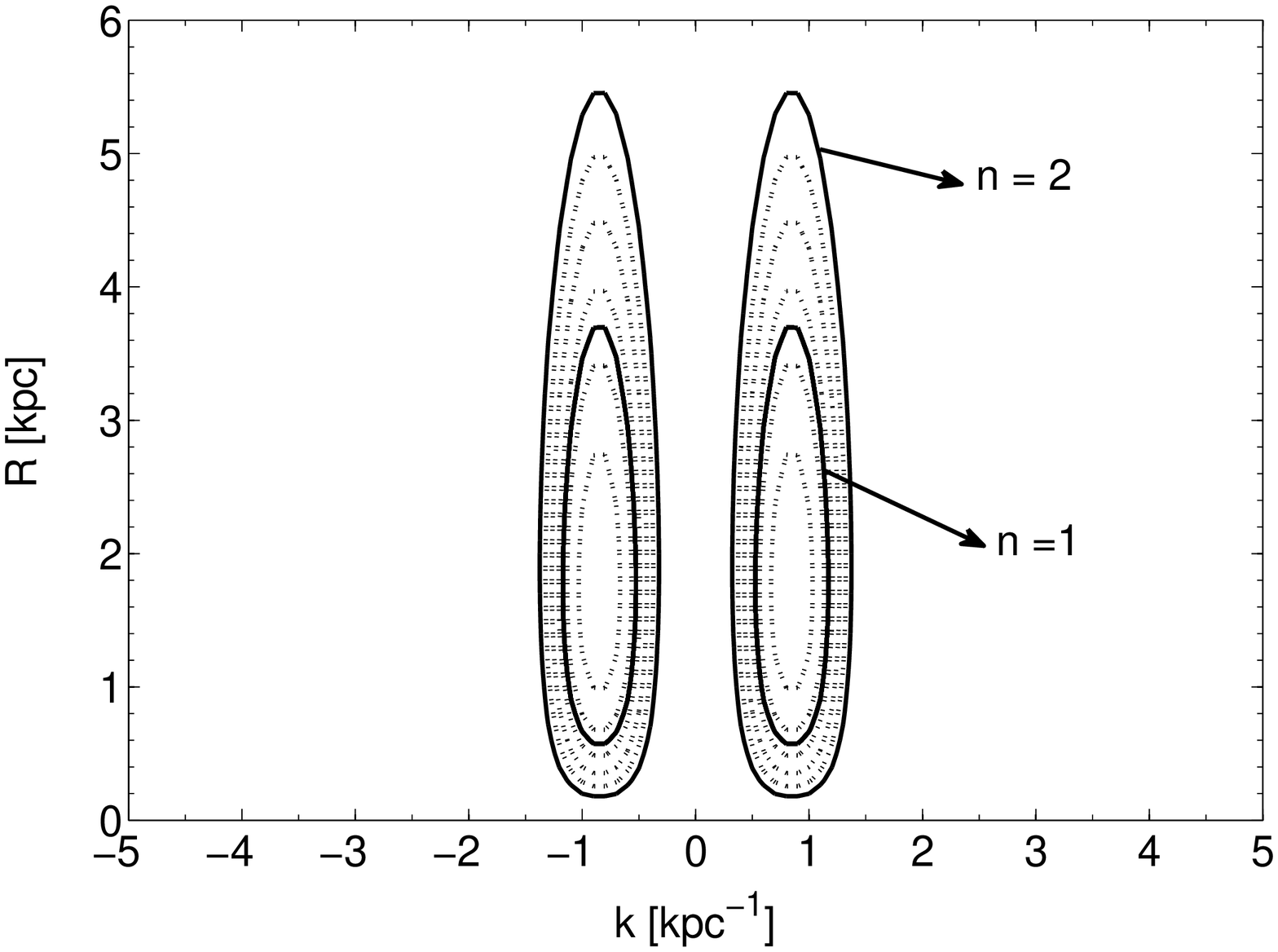}
\medskip
\includegraphics[height=2.5in,width=3.5in]{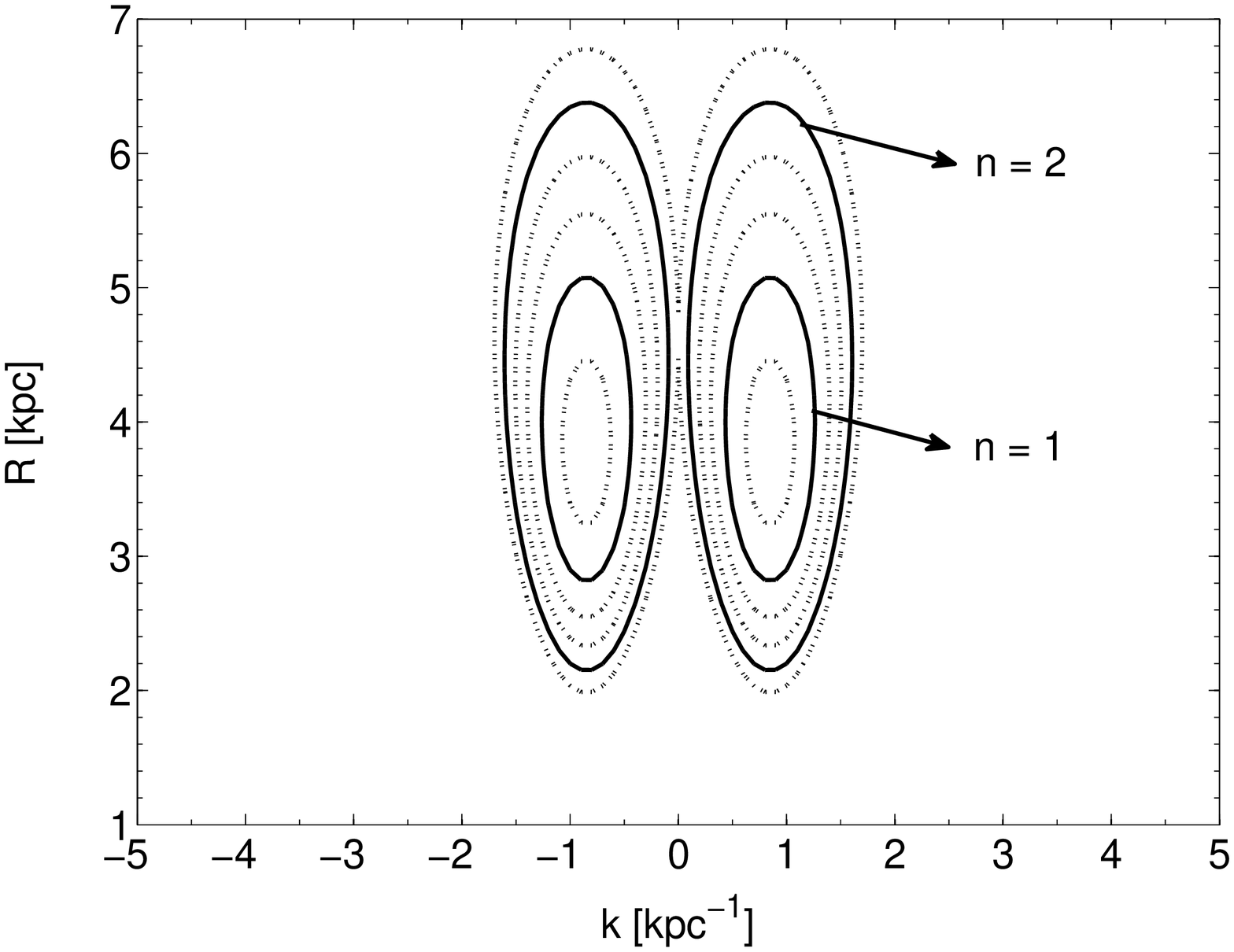}
\caption{Propagation diagrams (contours of constant $\Omega_{\rm p}$) for those pattern speeds which give closed loops of {\it{Type B}}. The input parameters used are for the LSB galaxy, UGC 7321. The top panel shows contours for disc-alone case where the range of $\Omega_{\rm p}$ varies from 3.8 km s$^{-1}$ kpc$^{-1}$ to 4.8 km s$^{-1}$ kpc$^{-1}$, at intervals of 0.2 and the bottom panel shows the contours for disc plus dark matter halo case where the range of $\Omega_{\rm p}$ varies from 3.6 km s$^{-1}$ kpc$^{-1}$ to 4.1 km s$^{-1}$ kpc$^{-1}$, at intervals of 0.1. The closed loops that correspond to the global modes for different models, are indicated by solid lines. $\Omega_{\rm p}$ value increases from the outer to the inner contours.}
\label{fig3}
\end{figure}

The results for the contours of {\it Type B} for both disc-alone and disc plus halo cases using the input parameters of UGC 7321 for different pattern speeds, are displayed in Fig.~\ref{fig3}.
We then applied the quantization condition (equation~(\ref{qcond})) to these contours to determine the quantum number $n$. The resulting values of $n$ for different pattern speeds that correspond to permitted global modes for both cases, are summarized in Table 1.
\begin{table}
\centering
\begin{minipage}{0.45\textwidth}
\caption{ Results for global modes for UGC~7321}
\begin{tabular}{ccccc}
\hline
$\Omega_{\rm p}$  &  $R_- $  & $R_+$ & $R_{\rm CR}$ & $n$ \\
(km s $^{-1}$ kpc$^{-1}$)&  (kpc) & (kpc) & (kpc) & \\
\hline
Disc-alone case :\\
\hline
4.5  &   0.5  & 3.8 & 7& 1\\
3.8  &  0.08  & 5.8  & 8& 2\\
\hline
Disc plus halo case :\\
\hline
4.0  &   2.8  & 5.1 & 33 & 1\\
3.7  &  2.1  & 6.4  & 36.5 & 2\\
\hline
\end{tabular}
\end{minipage}
\end{table} 
Note that in the statistical majority of galaxies, grand-design spiral structure should be quasi-stationary; and this can be possible if the dynamics of the disc is dominated by a single mode or by a small number of modes. Our finding, summarized in Table~1, agrees quite well with this point. Note that higher values of $n$ correspond to loops enclosing larger area, and for such loops $k_{-} \simeq 0$. Therefore, the smaller values of $n$ are more likely to satisfy the WKB approximation than the larger values of $n$.  

 However, we caution the reader that the modes that we have obtained are for a fluid disc (not collisionless), under the WKB approximation; and therefore the specific values may change if either of these assumptions is relaxed. Table 1 shows that the inclusion of dark matter halo has a negligible effect on the global spiral modes. The only change that we can observe from the Table 1 is that the inclusion of halo appreciably increases the extent of the forbidden region (determined from the difference of $R_{\rm CR}$ and $R_{+}$) and it only changes the specific values of pattern speed which correspond to different values of $n$. Consequently  the position of CR for different modes also changes. \citet{GJ14} showed that inclusion of dark matter halo suppresses the local, transient, swing-amplified spiral features almost completely and therefore it is somewhat puzzling why dark matter halo fails to make any impression on global modes permitted in a galactic disc, modelled as a fluid.

We argue that there are several reasons why global modes, though technically permitted, may not materialize in a LSB galaxy. First,  a galaxy encounter is known to be a possible mechanism to excite the global spiral modes in a galactic disc. For example, spiral structures of M51 and many other grand-design spirals with a companion have been modelled successfully \citep{TOMTOM72,BT87}. Since LSBs are observed to be isolated \citep{MO94}  or are located at the edges of voids \citep{ROSEM09}, they are less likely to undergo tidal encounters as compared to the HSB galaxies. Therefore, even if theoretically the modes are
permitted for the disc case, as well as the disc plus dark matter halo case for UGC 7321, in reality the triggering mechanism for exciting the global spiral modes may not operate. Second, the resulting pattern speed for the permitted global modes is small (Table 1). To excite these by interactions, a distant encounter is necessary which in turn will lead to a smaller amplitude. 
Third, even if the above permitted modes were to be triggered, the growth rate of the modes will depend on the surface density of the disc which is very low for the LSB galaxies, hence the amplitude of such modes is likely to be very small.
 Finally, we would like to mention that from the linear calculation done in this paper, we cannot fully comment on the effect of dark matter halo. 
A full N-body treatment which includes the above non-linear effects would be necessary in getting a comprehensive picture of the effect of dark matter halo on global spiral modes.

Note that in contrast, the lack of nearby neighbours does not prevent local modes being triggered in the discs of LSB galaxies \citep [studied by][]{GJ14}, since local, transient swing-amplified features can be generated by internal triggers such as a gas cloud or star formation \citep{Sellcal84,Tom90}.
In this paper, we have treated the stellar disc as a fluid for mathematical simplicity. At large values of $|k|$, the behaviour of the dispersion relation for a collisionless system is very different from that for a fluid system \citep{Raf01,GJ15}. In a future work, we will follow up this problem treating the galactic disc as a collisionless system.

\subsection {The Galaxy}
For comparison, we carried out a similar modal analysis for the Galaxy, for which the dark matter halo is known to be not a dominant component in the inner regions of the disc. The closed contours present in the propagation diagram for different pattern speeds for the Galaxy in both disc-alone and disc plus dark matter halo cases are shown in Fig~4.\\
\begin{figure}
\centering
\includegraphics[height=2.5in,width=3.5in]{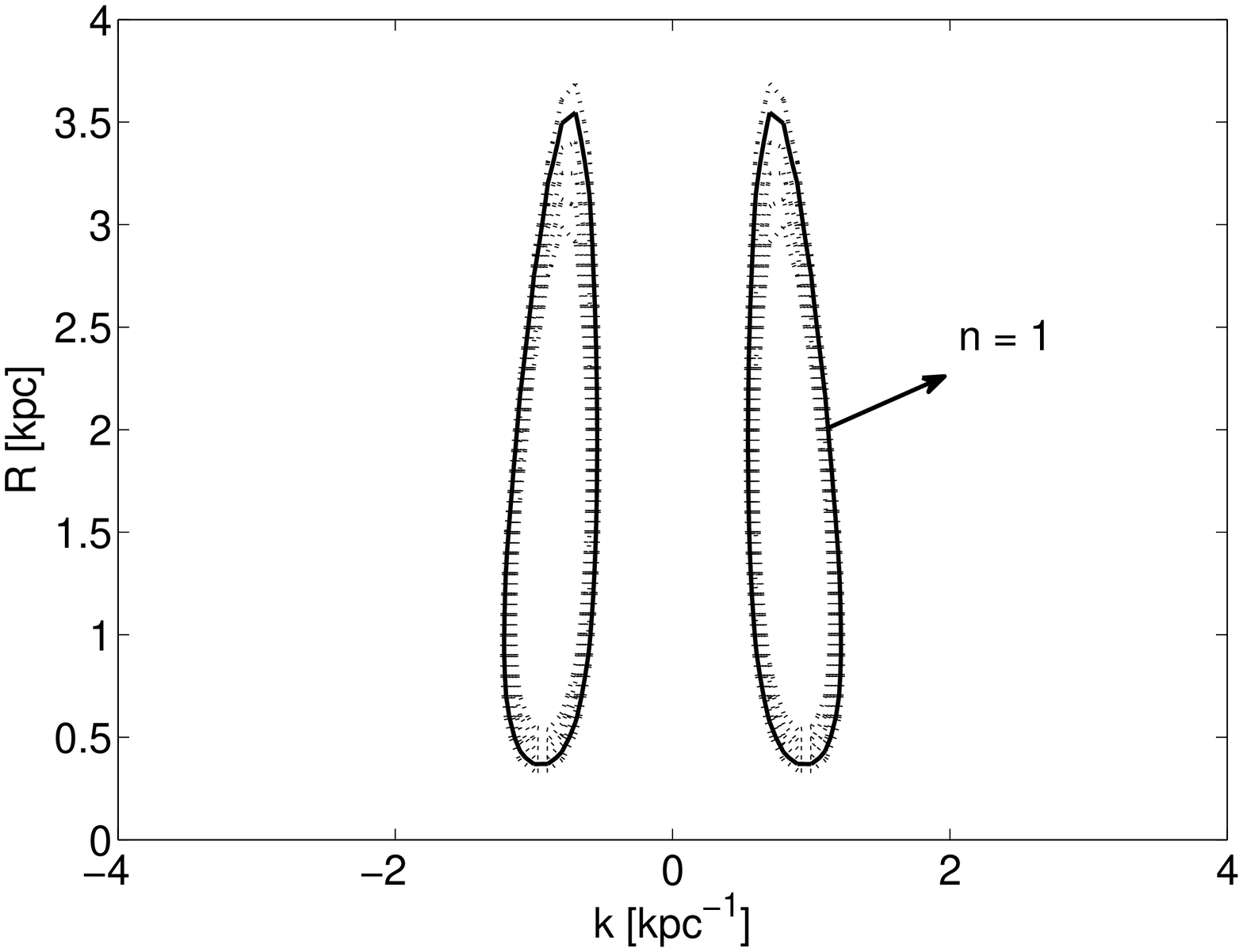}
\medskip
\includegraphics[height=2.5in,width=3.5in]{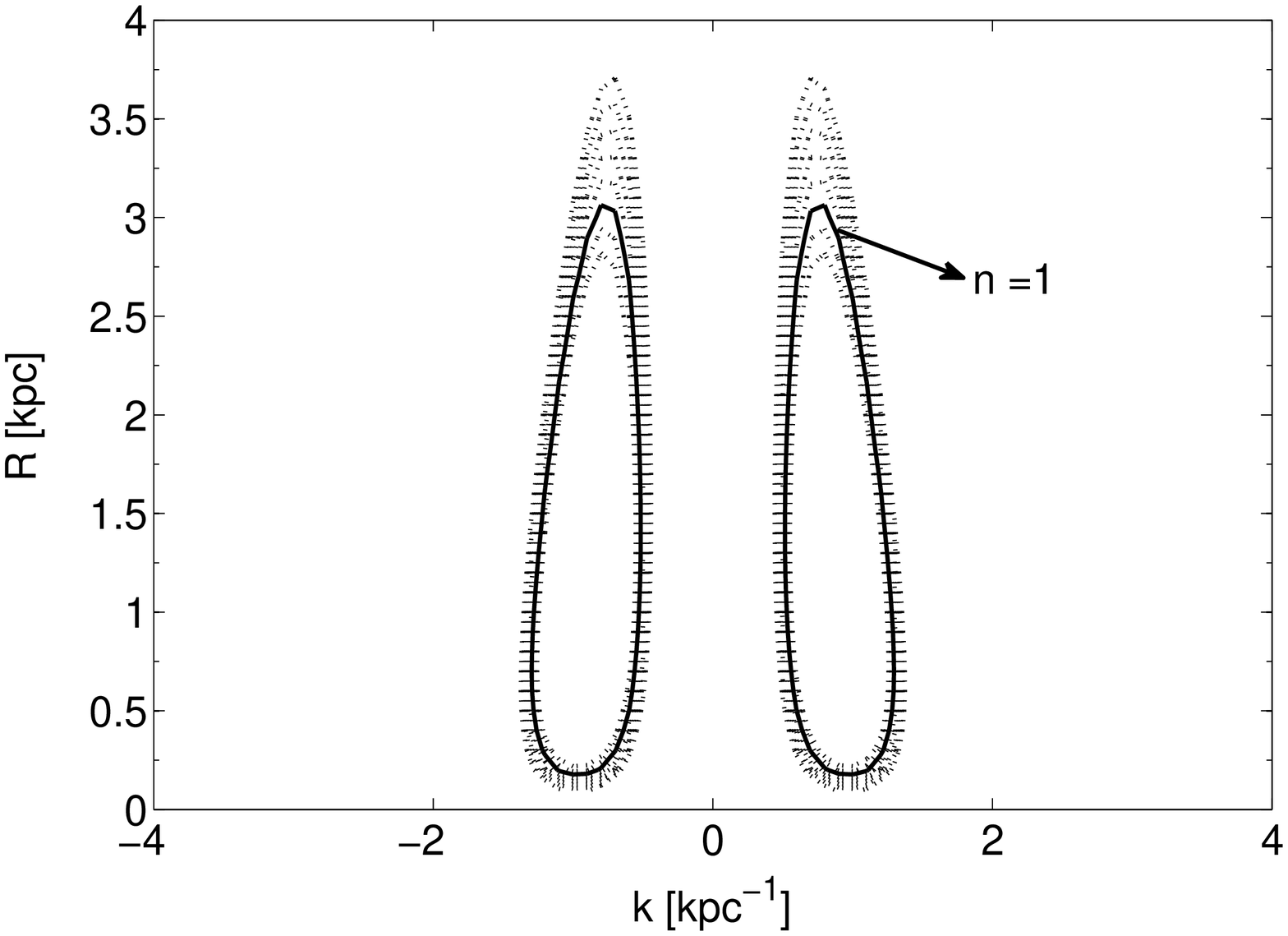}
\caption{Propagation diagrams (contours of constant $\Omega_{\rm p}$) for those pattern speeds which give closed loops of {\it{Type B}}. The input parameters used are for the Galaxy. The top panel shows contours for disc-alone case where the range of $\Omega_{\rm p}$ varies from 17 km s$^{-1}$ kpc$^{-1}$ to 18 km s$^{-1}$ kpc$^{-1}$, at intervals of 0.2 and the bottom panel shows the contours for disc plus dark matter halo case where the range of $\Omega_{\rm p}$ varies from 14 km s$^{-1}$ kpc$^{-1}$ to 15.4 km s$^{-1}$ kpc$^{-1}$, at intervals of 0.1. The closed loops that correspond to the global modes for different models, are indicated by solid lines. $\Omega_{\rm p}$ value increases from the outer to the inner contours.}
\end{figure}
 For a quantitative verification, following the same procedure as taken for UGC 7321, we calculated the possible modes that could be present in both models. These are summarized in Table~2, which shows that there is only a small quantitative change in the overall spiral mode properties when we include the dark matter halo in the system. But, note that, unlike the case of UGC 7321, on inclusion of dark matter halo, the extent of forbidden region has not increased appreciably. This is in the line of our expectation because within the solar radius, the dark matter halo contributes 50 per cent or less to the rotation curve \citep{KG91,Sack97}.

\begin{table}
\centering
 \begin{minipage}{.45 \textwidth}
\caption{Results for global modes for the Galaxy}
\begin{tabular}{ccccc}
$\Omega_{\rm p}$  &  $R_- $  & $R_+$ & $R_{\rm CR}$ & $n$ \\
(km s $^{-1}$ kpc$^{-1}$)&  (kpc) & (kpc) & (kpc) & \\
\hline
Disc-alone case :\\
\hline
17.2 & 0.3  & 3.6 & 8 & 1\\
\hline
Disc plus halo case:\\
\hline
15.2  & 0.2 & 3.0 & 13 & 1 \\
\hline
\end{tabular}
\end{minipage}
\end{table} 
 
\section{Discussion}
\subsection{Effect of Bulge}
At this point note that, for the Galaxy, the radial extent of the closed loops of ${\it Type ~ B}$ is in the very inner region (see figure 4) where the bulge component dominates, and the spiral structures are not believed to be present at those radii. So far in the models of both galaxies, we have not included bulge. Normally, in the early type spirals, bulge dominates in the inner regions, and the late-type (e.g. Scd type) galaxies have no significant bulge \citep[]{BM98}. Consequently, the exclusion of the bulge will lead to a significant underestimation of the rotation curve in the inner parts of the early-type galaxies.

In this section we have included the bulge component to the existing mass models, namely, disc-alone and disc plus dark matter cases for our Galaxy. On the other hand, UGC 7321 has no discernible bulge \citep{MAT99,MAT03}. It is a typical bulgeless galaxy \citep{Kau09}. So, we redid the global modal analysis, this time taking into account of the bulge component, for the Galaxy only. We adopt a Plummer-Kuzmin bulge model to derive the bulge contribution for the Galaxy.

In the spherical coordinates, the density profile of the bulge is given by the following formula (Binney \& Tremaine 1987):
\begin{equation}
\rho_{\rm bulge}= \frac{3 M_{\rm b}}{4 \pi R^3_{\rm b}}\Bigg(1+\frac{R^2}{R^2_{\rm b}}\Bigg)^{-5/2}\,
\end{equation}
where $R_{\rm b}$ is  the  bulge  scale  length  and $M_{\rm b}$ is  the  total  bulge mass.\\
The corresponding potential in the cylindrical coordinates ($R$, $\phi$, $z$) is given as
\begin{equation}
\Phi_{\rm bulge} (R, z)=-\frac{G M_{\rm b}}{R_{\rm b}}{\Bigg(1+\frac{R^2+z^2}{R^2_{\rm b}}\Bigg)^{-1/2}}
\end{equation}
Corresponding rotational frequency $\Omega_{\rm bulge}$ and the epicyclic frequency $\kappa_{\rm bulge}$ in the mid-plane ($z$ = 0), are
\begin{equation}
\Omega^2_{\rm bulge}=\frac{G M_{\rm b}}{R^3_{\rm b}}\Bigg(1+\frac{R^2}{R^2_{\rm b}}\Bigg)^{-3/2}\,
\end{equation}
and,
\begin{equation}
\kappa^2_{\rm bulge}=\frac{G M_{\rm b}}{R^3_{\rm b}}\Bigg[4\Bigg(1+\frac{R^2}{R^2_{\rm b}}\Bigg)^{-3/2}\\
-3\Bigg(\frac{R}{R_{\rm b}}\Bigg)^2\Bigg(1+\frac{R^2}{R^2_{\rm b}}\Bigg)^{-5/2}\Bigg]
\end{equation}
Therefore, while taking the bulge component into account, these terms $\kappa^2_{\rm bulge}$ and $\Omega^2_{\rm bulge}$ will be added in R. H. S. of equation (1).
For the parameters of the bulge component, we have used $R_{\rm b}$ = 2.5 kpc and $M_{\rm b}$ = 2.8 $\times$ $10^8$ $M_{\odot}$ \citep{BLUM95}.

We did a global mode analysis, similar to that in \S~ 3, for the disc plus bulge case, and then for the disc plus bulge plus dark matter halo case. As explained earlier, we considered only closed loops of $\it{Type ~ B}$ for both cases. Then the relevant quantization condition is applied to obtain the principle quantum number $n$. The results are summarized in Table.3.

\begin{table}
\centering
 \begin{minipage}{.45 \textwidth}
\caption{Results for global modes for the Galaxy (including bulge)}
\begin{tabular}{ccccc}
$\Omega_{\rm p}$  &  $R_- $  & $R_+$ & $R_{\rm CR}$ & $n$ \\
(km s $^{-1}$ kpc$^{-1}$)&  (kpc) & (kpc) & (kpc) & \\
\hline
Disc plus bulge case :\\
\hline
18.3 & 1.8  & 4.5 & 9.7 & 1\\
15.1 & 1.1  & 6.3 & 11.3 & 2\\
\hline
Disc plus bulge plus halo case:\\
\hline
17.1 & 1.6  & 4.2 & 13.3 & 1\\
14.0 & 0.9  & 5.8 & 16 & 2\\
\hline
\end{tabular}
\end{minipage}
\end{table} 
In both models where bulge is included, two modes are present while models without bulge gave only one mode (see Table 2. and 3.). More importantly, the extent of global spiral arms (indicated by $R_{-}$ and $R_{+}$) shows an extended range and closer to the range where arms are observed \citep{BM98}. Note that the resulting pattern speed for $n=1$ is close to the observed value for the Galaxy \citep{Sie12}.
\subsection {Other issues}
In this subsection, we would like to mention  a few other issues regarding this work. First, in this work, we have not considered the low dispersion component in the disc, namely gas. In reality, a spiral galaxy contains a finite amount of gas. The role of gas has been studied in several dynamical issues, e. g. in the stability of local axisymmetric perturbations \citep{JS84a,JS84b, BR88,Raf01}, local non-axisymmetric perturbations \citep{Jog92}, and in the radial group transport \citep{GJ15} etc. Here in this paper, our main aim was to address the role of only dark matter halo on the grand-design spiral structure, hence we excluded gas from this formalism. This ensures that whatever change we see in these models, they are purely due to the dark matter halo. Besides, the superthin LSB galaxy UGC 7321 considered here has small gas content as compared to that of normal HSB galaxies \citep{UM03}. 
A full investigation of role of DM halo in a system with gas and stars coupled via gravitation will be followed in a future work.

Another point to note is that, this calculation is based on the linear perturbation theory, i. e. all the non-linear effects are neglected. Interestingly a recent study by \citet{Don13} using high resolution N-body simulations shows that the nonlinear effect can significantly modify the origin and persistence scenario of flocculent spiral features. So, it is worth checking the nonlinear effect in this context with gas treated on an equal footing with stars.

\section{Conclusions}
We have studied the existence of global modes in spiral galaxies in the WKB limit using the Bohr-Sommerfeld quantization condition. This approach has been used for the first time to study the effect of dark matter halo on the grand-design spiral structure in a galactic disc. Using the input parameters of a typical superthin LSB galaxy  UGC 7321, we found that in both disc-alone and disc plus halo cases, global modes are permitted. While the small-scale spiral structure is suppressed by the dark matter halo \citep{GJ14}, the halo does not have a significant effect on the global spiral modes. We argue that the tidal interactions causing global spiral modes are less likely to occur in LSBs as compared to the HSB counterparts. Thus even though the model for UGC 7321 including dark matter halo permits the existence of global spiral modes, in reality they are not likely to materialize, since these galaxies are isolated and hence may not experience tidal forces due to galaxy encounters.
Also, we carried out a similar analysis for our Galaxy. We found that, for our Galaxy, inclusion of the dark matter halo does not affect the existence of global spiral modes.
Note, however, that these results are obtained while treating the galactic disc as a fluid. These results may get modified when the galactic disc is treated as a collisionless system. This problem of investigating the role of dark matter halo on large-scale spiral arms in a collisionless galactic disc will be followed up in a future paper.

\bigskip
\noindent {\bf Acknowledgements:} We thank the anonymous referee for constructive comments which have greatly improved the presentation in the paper. C.J. would like to thank the DST, Government of India for support via a 
J.C. Bose fellowship (SB/S2/JCB-31/2014).

\end{document}